\documentclass[twocolumn,english,english,nobibnotes,nofootinbib]{revtex4-1}
\usepackage[T1]{fontenc}
\usepackage[latin9]{inputenc}
\usepackage{mathtools}
\usepackage{amsmath}
\usepackage{amssymb}
\usepackage{graphicx}
\usepackage{esint}

\makeatletter

\newcommand{\lyxdot}{.}

\usepackage[english]{babel}
\usepackage{braket}

\usepackage{tikz}%

\usepackage{babel}

\usepackage{babel}

\makeatother

\usepackage{babel}
\begin{document}

\title{Imaging snake orbits at graphene n-p junctions}

\author{K. Kolasi\'{n}ski }

\affiliation{AGH University of Science and Technology, Faculty of Physics and
Applied Computer Science,\\
 al. Mickiewicza 30, 30-059 Kraków, Poland}

\author{A. Mre\'{n}ca-Kolasi\'{n}ska }

\affiliation{AGH University of Science and Technology, Faculty of Physics and
Applied Computer Science,\\
 al. Mickiewicza 30, 30-059 Kraków, Poland}

\author{B. Szafran}

\affiliation{AGH University of Science and Technology, Faculty of Physics and
Applied Computer Science,\\
 al. Mickiewicza 30, 30-059 Kraków, Poland}

\begin{abstract}
We consider conductance mapping of the snake-orbits confined along the n-p junction defined in graphene by the electrostatic doping
in the quantum Hall regime. %The conductance matrix elements are evaluated in the Landauer approach for a four-terminal device. 
We explain the periodicity of conductance oscillations
at the magnetic field and the Fermi energy scales by the properties of the n-p junction as a conducting channel. 
We evaluate the conductance maps for a floating gate scanning the surface of the device. In the quantum Hall conditions
the currents flow near the edges of the sample and along the n-p junction. The conductance mapping resolves only the n-p junction and not the edges.
The conductance oscillations  along the junction are found in the maps with periodicity related to the cyclotron orbits of the
scattering current. Stronger probe potentials provide support to localized resonances at one of the sides of the junction with current loops
that interfere with the n-p junction currents. The interference results in a series of narrow lines parallel to the junction with positions
that strongly depend on the magnetic field through the Aharonov-Bohm effect. The consequences of a limited transparency of finite width n-p junctions  are also discussed.
\end{abstract}
\maketitle

\section{Introduction}

The gapless band structure of graphene allows for electrical doping
with formation of n-type and p-type regions defined by external potentials
\cite{CastroNeto2009Rev}. With the electron mean free path \cite{mfp,cbbt4}
of several $\mu$m, the n-p junctions \cite{Ghosh2008,Levitov2007,tworzy}
in graphene are attractive playground for studies of electron optics
\cite{vl,szaczen,Rickhaus2013,tacza,cbeo,richterliu} implemented
in the solid state. In particular, the n-p junctions in the quantum
Hall conditions \cite{Williams2007,Abanin2007,Huard2007,Barbaros2007}
form waveguides \cite{wi11} for electron currents, which in the semi-classical
picture go along snake orbits \cite{snakepee,snak3,snak33,snak5,snak6,snak7,snakepee2,Milovanovic2014,Rickhaus2015,tacza,Mele2015SnakeSym,Zarenia2013,Oroszlany2008}
formed by inversion \cite{snakemu} of the orientation of the Lorentz
force \cite{Szafran2005} with the carriers passing across the junction
from the conductance to the valence band.

In this work we consider the possibility of mapping the snake orbits
confined along the n-p junction using the scanning gate microscopy
\cite{Topinka2000,Topinka2001,Kozikov2015,Jura2009,Sellier2011,Morikawa2015,Bhandari2016}
(SGM). The SGM microscopy \cite{Topinka2000,Topinka2001,Kozikov2015,Jura2009,Sellier2011,Morikawa2015,Bhandari2016}
with a charged tip of an atomic force microscope floating above the
sample allows one to probe the quantum transport properties of devices
with a spatial resolution. Cyclotron and skipping orbits -- which
are related to the snake orbits by the magnetic deflection -- have
already been experimentally resolved \cite{Morikawa2015,Bhandari2016}
for magnetic focusing \cite{ma13,temf1,Stegmann2015,richterliu,Milovanovic2014,Taychatanapat2013}
of electron currents in unipolar graphene sheets.

For the purpose of the present work we consider a four-terminal cross
junction -- a geometry of a quantum Hall bar studied previously both
by experiment \cite{cbbt,cbbt2,cbbt3,cbbt4,cbeo} and theory \cite{Milovanovic2014,snakepee}
of ballistic transport in graphene. We determine the transport properties
of the n-p junction defined within the sample using the atomistic
tight-binding approach. We find the characteristic conductance oscillations
\cite{Milovanovic2014,Rickhaus2015,tacza,Milovanovic2014,Stegmann2015}
as a function of the magnetic field that are identified with formation
of snake orbits. The experimental conductance oscillations can be
exactly reproduced by a coherent quantum transport simulation as shown
in Ref. \cite{Rickhaus2015}. In this work we explain the periodicity
of the conductance oscillations by the details of the dispersion relation
of the n-p junction waveguide \cite{cohn,szaczen} as due to superposition
of the junction modes producing scattering density oscillations of
the largest wavelength. A perfect agreement with the results of the
quantum transport simulation is found in the entire quantum Hall regime.

We demonstrate that the potential of the scanning probe produces variation
of the sample conductance but only when the probe floats above the
n-p junction. The probe deflects the electron paths changing the destination
terminal of the electron currents and thus affecting the conductance.
Outside the junction the sample does not react to the probe as the
backscattering is suppressed in the quantum Hall conditions. Although,
the electron paths are not as clearly resolved as for the magnetically
focused trajectories \cite{Morikawa2015,Bhandari2016}, the period
of the conductance oscillations along the junction is close to the
length of the snake orbit period. For stronger tip potentials series
of resonances are found on the lines parallel to the junction -- but
only on one of its sides -- where the tip potential supports formation
of the quasi-bound states. For these resonances a current loop is
found around the probe which interferes with the wave function flow
along the junction waveguide.

\section{Theory}
\subsection{Model Hamiltonian}
We consider a four-terminal cross structure which is depicted in Fig.
\ref{fig:sketch}(a). %The source and drain gates are labelled%with $L_{1}$ and $L_{2}$. Two voltage probes ($L_{3}$ and $L_{4}$)%measure the voltage drop across the n-p junction (see dashed lines%in (a)).
We use the tight-binding Hamiltonian 
\begin{equation}
\boldsymbol{H}=\sum_{k}U_{k}\left(\mathbf{r}_{k}\right)\boldsymbol{c}_{k}^{\dagger}\boldsymbol{c}_{k}+\sum_{\left\langle i,j\right\rangle }t_{ij}\boldsymbol{c}_{i}^{\dagger}\boldsymbol{c}_{j}+h.c,\label{eq:H}
\end{equation}
where the second summation denotes the nearest neighbor pairs and $U_{k}\left(\mathbf{r}_{k}\right)$
is the external on-site potential energy on the $k$-th site in the
lattice. The magnetic field is taken into account by the Peierl's
substitution 
\[
t_{ij}=t\exp\left(\frac{2\pi ei}{h}\right)\int_{\mathbf{r}_{i}}^{\mathbf{r}_{j}}\mathbf{A}\cdot\mathbf{dl}
\]
with the hoping energy $ T=-$2.7eV. % and carbon-carbon distance $a_{\mathrm{CC}}=1.42\AA$.
In this paper we consider the magnetic field perpendicular to the
graphene surface $\mathbf{B}=\left(0,0,B\right)$ and use the Landau
gauge $\mathbf{A}=\left(-By,0,0\right)$. %The scattering region  contains%190000 carbon atoms in total. 
In order to model samples of linear size of about 300 nm we apply the
scaling method proposed in Ref. \cite{Liu2015Graphene}
with the scaling factor of $10$. 
 %\textcolor{red}{Lepszy opis próbki,%PMMA, itd \.{z}e jest ekranowanie od dolnego leadu i \.{z}eby zmienia\'{c}%Ef w ca\l{}ej próbce trzeba by by\l{}o zmienia\'{c} jednocze\'{s}nie%back gate is bottom Jednak nas to nie zabardzo interesuje bo my skupiamy%sie tylko na ustalonej wartosci potencja\l{}ów i zmieniamy pole B?}%Such technique was successfully applied in recent work and allows%to approach the graphene surface with the tip of the SGM on entire%sample {[}arXiv:1608.07503{]}. \textcolor{red}{Chocia\.{z} Rickhaus%\cite{Rickhaus2015} te\.{z} mia\l{} tak\k{a} próbke gdzie da\l{}o%by si\k{e} podej\'{s}\'{c} od góry i mieli super wyniki wi\k{e}c mo\.{z}e%nie pisa\'{c} o tym co Heun zrobi\l{}???.}

We consider a system in which the n-p junction is formed by external potentials induced
by the gate electrodes along the diagonal of the cross-junction. 
The potential profile of the junction is modeled with an analytical
formula 
\begin{equation}
U_{PNJ}(x)=\frac{eV_{PNJ}}{e^{-x'/S_{\mathrm{m}}}+1},\label{eq:Vpnj}
\end{equation}
where $x'$ axis coincides with the $y=x$ line with the origin at  the diagonal of the cross-junction [see the dashed blue line in Fig. 1(a,b)]. In Eq. (\refeq{eq:Vpnj}) $eV_{PNJ}$ is the potential energy variation across the junction, and
$S_{\mathrm{m}}$ controls the width of the n-p interface.  % which controls the spatial width of the%pn interface. % The abrupt junction for $S_{\mathrm{m}}=0$ we%as in Ref. \cite{Milovanovic2014}),
% By changing the value of $V_{PNJ}$%parameter one changes the electron density in the right region. When
The potential \eqref{eq:Vpnj} for $eV_{PNJ}>E_{\mathrm{F}}$ induces the n-p junction 
with the p- type conductivity in the upper-right part of the device of Fig. 1(a).
In this paper we follow the choice of Ref. \cite{Rickhaus2015} and
restrict our considerations to the symmetric case when the carrier
densities are the same in the n- and p-type regions, i.e. for $eV_{PNJ}=2E_{\mathrm{F}}$.

In order to simulate the SGM mapping we use the Lorentzian approximation
for the tip-induced potential 
\[
U_{\mathrm{tip}}\left(\mathbf{r};\mathbf{r}_{\mathrm{tip}}\right)=\frac{d_{\mathrm{tip}}^{2}V_{\mathrm{tip}}}{\left|\mathbf{r}-\mathbf{r}_{\mathrm{tip}}\right|^{2}+d_{\mathrm{tip}}^{2}},
\]
where $\mathbf{r}_{\mathrm{tip}}$ is the position of the tip, $d_{\mathrm{tip}}$
-- the Lorentzian width and $V_{\mathrm{tip}}$ -- the tip induced
amplitude. The tip induced potential is controlled by external voltages,
and its width is close to the distance between the tip and the sheet
that confines the electron gas \cite{szafranDFT2011}. Here, we choose
 $d_{\mathrm{tip}}=25$ nm.  The potential energy $U_{\mathrm{tip}}\left(\mathbf{r};\mathbf{r}_{\mathrm{tip}}\right)$
enters the on-site term of the Hamiltonian \eqref{eq:H}, i.e., $U=U_{\mathrm{tip}}+U_{PNJ}$.

\subsection{Conductance}
In order to evaluate the transport properties of the device we use
the Landauer-B\"uttiker approach together with the wave function matching
method \cite{Sorensen2009,Kolasinski2016} which requires a numerical
solution of the scattering problem. A low temperature 
$\sim0$K and a source-drain bias within the linear response regime are assumed.

\begin{figure}
\begin{centering}
\includegraphics[width=1\columnwidth]{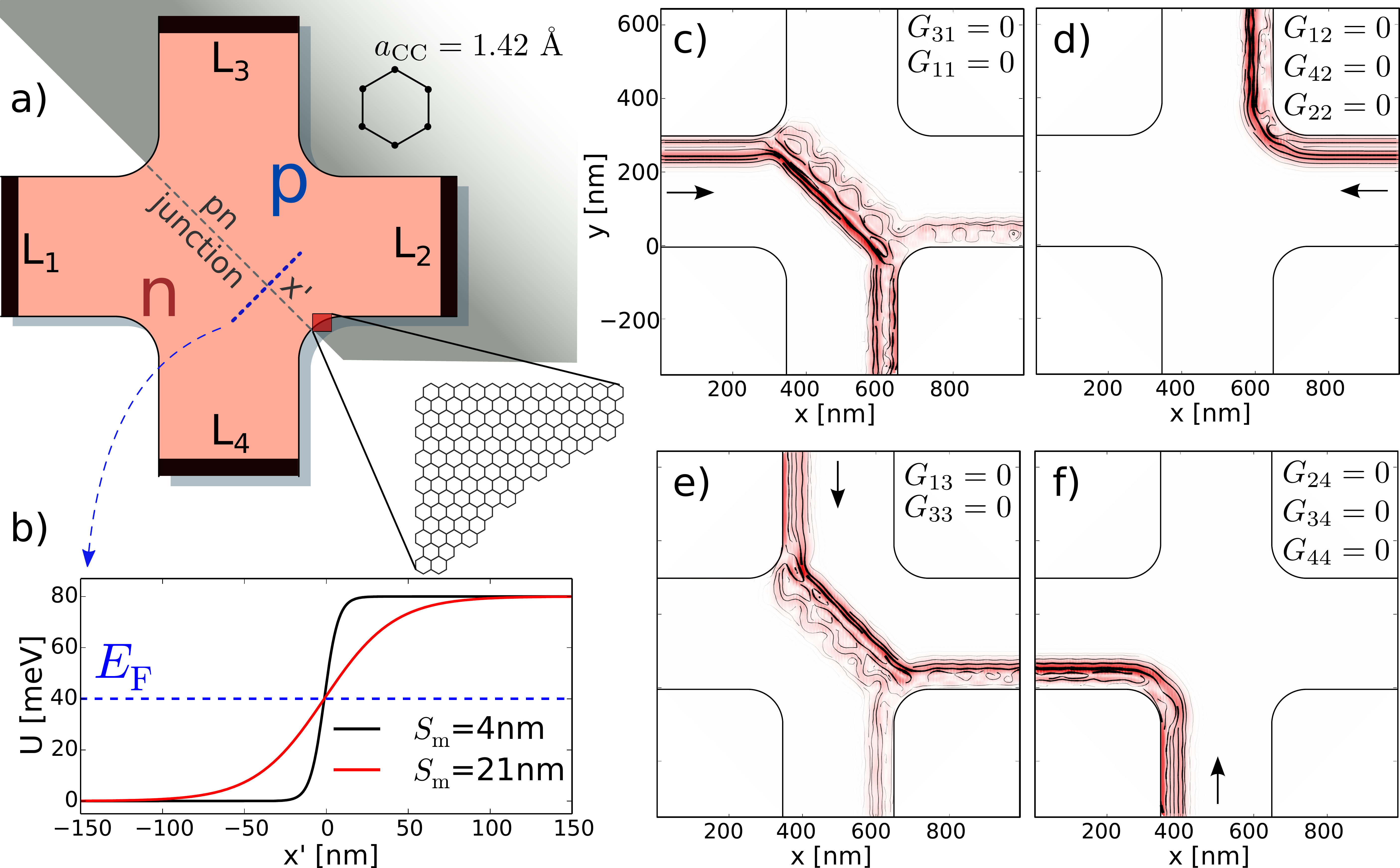} 
\par\end{centering}

\caption{\label{fig:sketch}a) The sketch of the four terminal device.
The source and drain leads are denoted by $L_{1}$ and $L_{2}$
labels, respectively. $L_{3}$ and $L_{4}$ leads are voltage probes.
The gray dashed line denotes the n-p junction induced by an external potential. (b) The potential profiles along the blue line in (a) for $S_{\mathrm{m}}=4$ nm and
$S_{\mathrm{m}}=21$ nm modeled with Eq. \eqref{eq:Vpnj}.
The n-region is located to the left from the n-p interface. (c-f) The probability current maps obtained for electron
incoming from each of the leads $L_{i}$ (incident direction is denoted by
the black arrow). The maps are computed for $B=1.3$ T, $E_{\mathrm{F}}=eV_{PNJ}/2=40$ meV. }
\end{figure}

The probability currents obtained for $B=1.3$ T, $E_{\mathrm{F}}=40$meV
and electron incident from different leads are shown in Fig. \ref{fig:sketch}(c-f),
with snake-like trajectories in Figs. \ref{fig:sketch}(c) and \ref{fig:sketch}(e).
Note, that already for this magnetic field value i) the current flows only near the edges \cite{Haug1993EdgeState}
and along the n-p junction and ii) backscattering to
the input lead is absent.

The conductance from lead p to $q$ denoted as $G_{qp}$ is computed
from the Landauer formula 
\begin{equation}
G_{qp}=G_{0}\sum_{m,n}\left|t_{p,m}^{{q,n}}\right|^{2},\label{eq:Gpq}
\end{equation}
where $ T_{p,m}^{{q,n}}$ is the scattering transmission amplitude
that electron entering the device at lead p in mode $m$ will leave
the device at lead $q$ in mode n. The summation runs over all
propagating modes in the leads and $G_{0}$ stands for the conductance quantum $G_{0}\equiv\frac{2e^{2}}{h}$. For more details about
the applied computational method  see e.g. \cite{Sorensen2009,Kolasinski2016,Zwierzycki2008}.

%\subsection{High magnetic field limit}

The conductance matrix $\mathbf{G}$ elements \cite{DattaBook} for
the four-terminal device are determined from solution of the scattering
problem according to Eq. \eqref{eq:Gpq} %\[%\mathbf{G}=\left(\begin{array}{cccc}%g_{11} & -G_{21} & -G_{31} & -G_{41}\\%-G_{12} & g_{22} & -G_{32} & -G_{42}\\%-G_{13} & -G_{23} & g_{33} & -G_{43}\\%-G_{14} & -G_{24} & -G_{34} & g_{44}%\end{array}\right)%\]%with $g_{ii}=\sum_{j\neq i}G_{ji}.$

\[
\mathbf{G}=\left(\begin{array}{cccc}
g_{11} & -G_{12} & -G_{13} & -G_{14}\\
-G_{21} & g_{22} & -G_{23} & -G_{24}\\
-G_{31} & -G_{32} & g_{33} & -G_{34}\\
-G_{41} & -G_{42} & -G_{43} & g_{44}
\end{array}\right)
\]
with $g_{ii}=\sum_{j\neq i}G_{ij}.$

In the quantum Hall conditions some elements of this matrix are
zero (see Figs. \ref{fig:sketch}(c-f)). In our device the quantum
Hall edge transport appears for $B\gtrsim0.8$ T, and then the conductance
matrix acquires the form

%\[%\mathbf{G}=\left(\begin{array}{cccc}%G_{21}+G_{41} & -G_{21} & 0 & -G_{41}\\%0 & G_{32} & -G_{32} & 0\\%0 & -G_{23} & G_{23}+G_{43} & -G_{43}\\%-G_{14} & 0 & 0 & G_{14}%\end{array}\right)%\]
\[
\mathbf{G}=\left(\begin{array}{cccc}
G_{14} & 0 & 0 & -G_{14}\\
-G_{21} & G_{21}+G_{23} & -G_{23} & 0\\
0 & -G_{32} & G_{32} & 0\\
-G_{41} & 0 & -G_{43} & G_{41}+G_{43}
\end{array}\right)
\]
Additionally, we have $G_{23}+G_{43}=G_{32}=\nu_{\mathrm{p}}$ and
$G_{21}+G_{41}=G_{14}=\nu_{\mathrm{n}}$ with $\nu_{\mathrm{p/n}}$
being the spin degenerated filling factors in the p/n-regions.
In the linear transport conditions the current in each of the leads
is given by $\mathbf{I}=\mathbf{G}\mathbf{V}$ for a given bias. We
choose $L_{1}$ and $L_{2}$ to be a source-drain electrodes -- with
the source $V_{1}=V_{\mathrm{S}}$,  and the drain $V_{2}=V_{\mathrm{D}}$
potentials, respectively. Terminals $L_3$ and $L_4$ are used as voltage
probes, which amounts in $I_{3}=I_{4}=0$ and so $I_{1}=-I_{2}$ (the
plus sign stands for the current that enters the device). The condition
$I_{3}=I_{4}=0$  immediately implies $V_{3}=V_{D}$. This fact can
be also deduced from Fig. 1(d) -- the current from the drain terminal
passes to $L_{3}$ without scattering. From the form of the second
row of the ${\bf G}$ matrix, we have $I_{2}=-G_{21}V_{S}+\left(G_{21}+G_{23}\right)V_{D}-G_{23}V_{D}=G_{21}(V_{D}-V_{S})$.
Hence, in the quantum Hall conditions the current flow is determined
uniquely by $G_{21}$ matrix element, which is studied in detail below.

\section{Results}
\subsection{Conductance oscillations due to snake orbits}

\begin{figure}
\begin{centering}
\includegraphics[width=1\columnwidth]{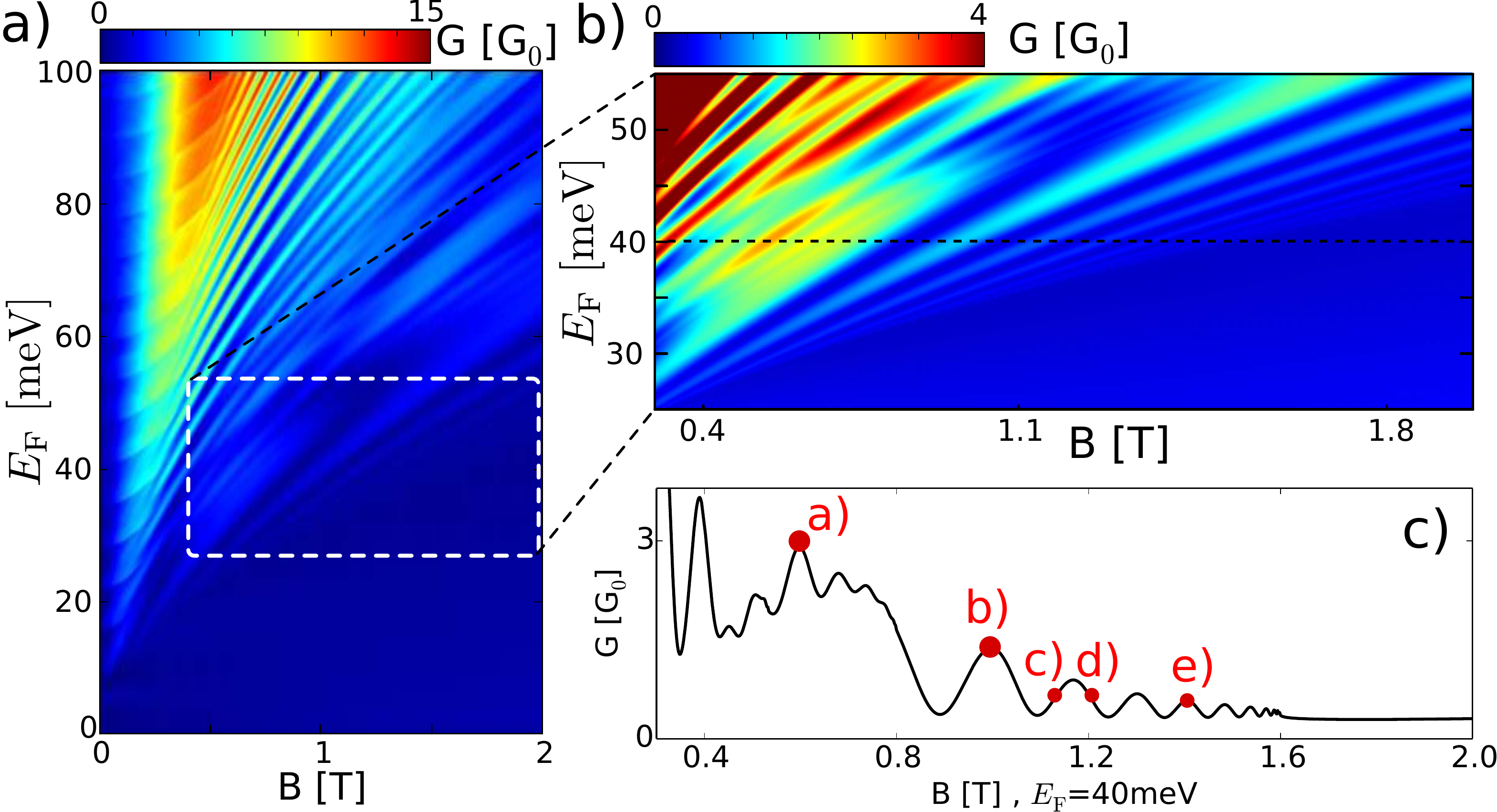} 
\par\end{centering}

\caption{\label{fig:a1}(a) $G_{21}$ conductance as a function of magnetic field
$B$ and Fermi energy $E_{\mathrm{F}}=eV_{PNJ}/2$.   (b) Zoom
of the white rectangle in (a).
(c) Cross section along the dashed line in (b). The red
labels correspond to the SGM images in Figs. \ref{fig:smg}(a-e).}
\end{figure}

Figure \ref{fig:a1}(a) shows the $G=G_{21}$ conductance as a function
of the magnetic field and the Fermi energy. The result contains an oscillatory pattern -- marked with the dashed rectangle -- 
similar to the one found \cite{Rickhaus2015} experimentally and identified
with formation of snake orbits along the n-p junction. The snake
states oscillations are visible in a quite large range of both  $E_{\mathrm{F}}$
and $B$. For further studies we fix the value of Fermi energy $E_{\mathrm{F}}=40$
meV and analyze the results as a function of magnetic field. The zoomed
fragment of Fig. \ref{fig:a1}(a) near $E_{\mathrm{F}}=40$ meV is
depicted in Fig. \ref{fig:a1}(b) and the cross section of the plot
for $E_{\mathrm{F}}=40$ meV is given in Fig. \ref{fig:a1}(c).

\begin{figure}[!h]
\begin{centering}
\includegraphics[width=1\columnwidth]{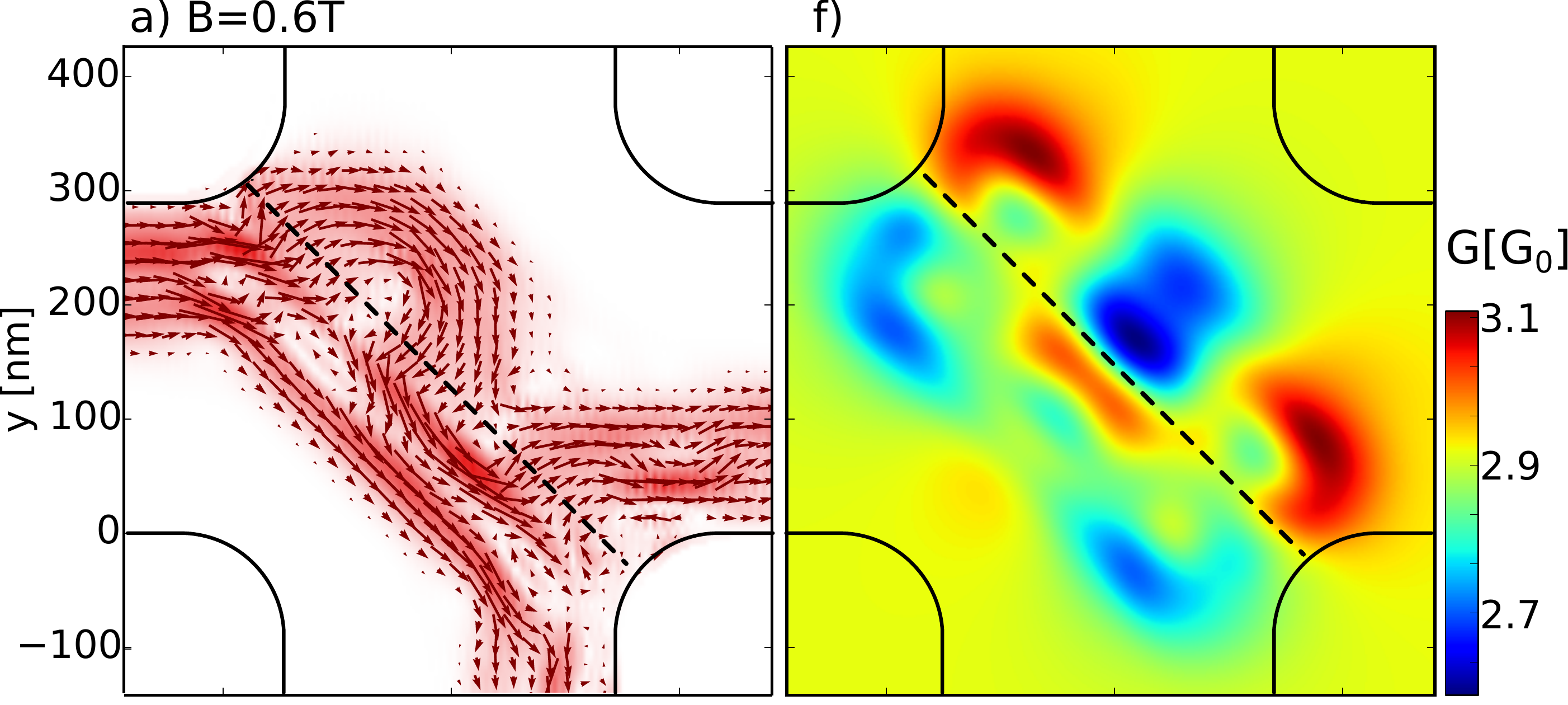} 
\par\end{centering}

\begin{centering}
\includegraphics[width=1\columnwidth]{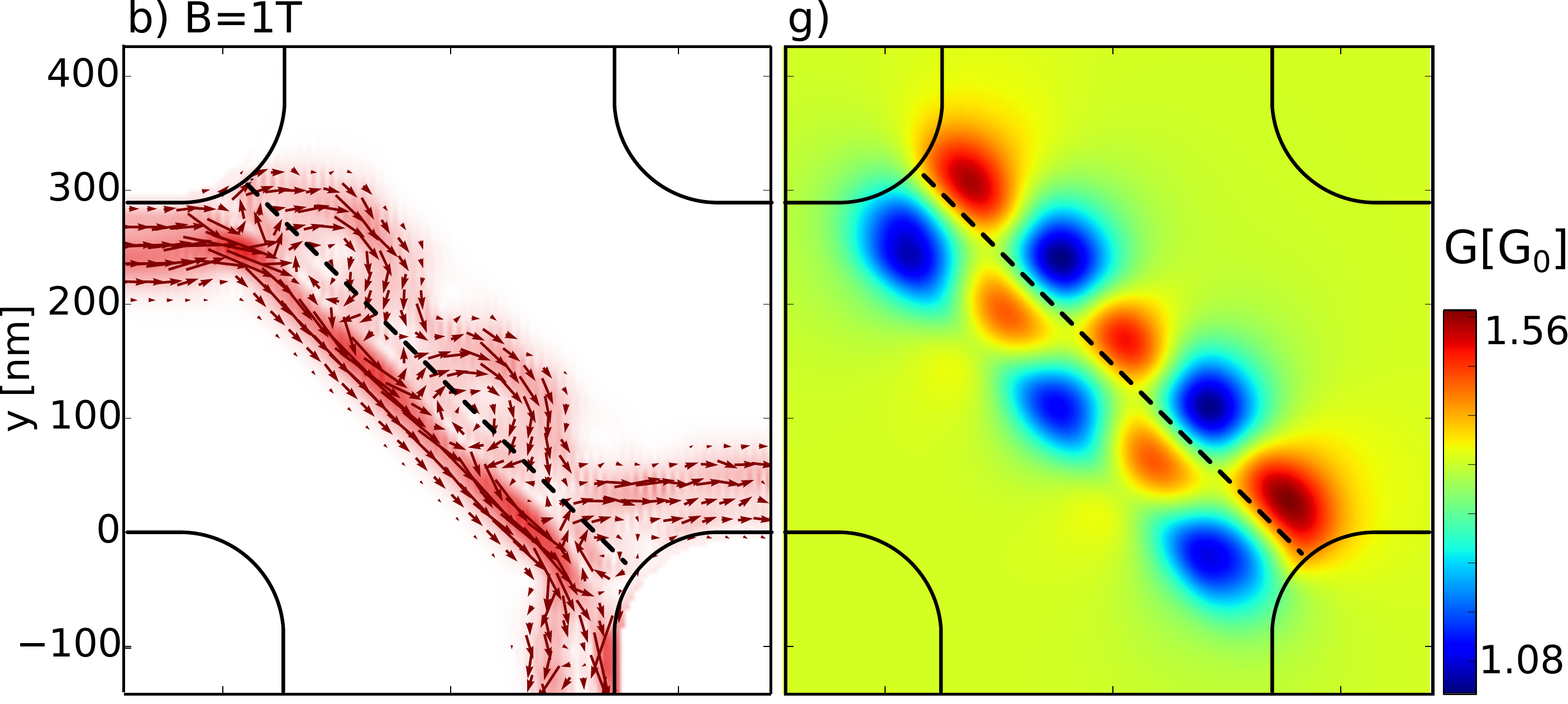} 
\par\end{centering}

\begin{centering}
\includegraphics[width=1\columnwidth]{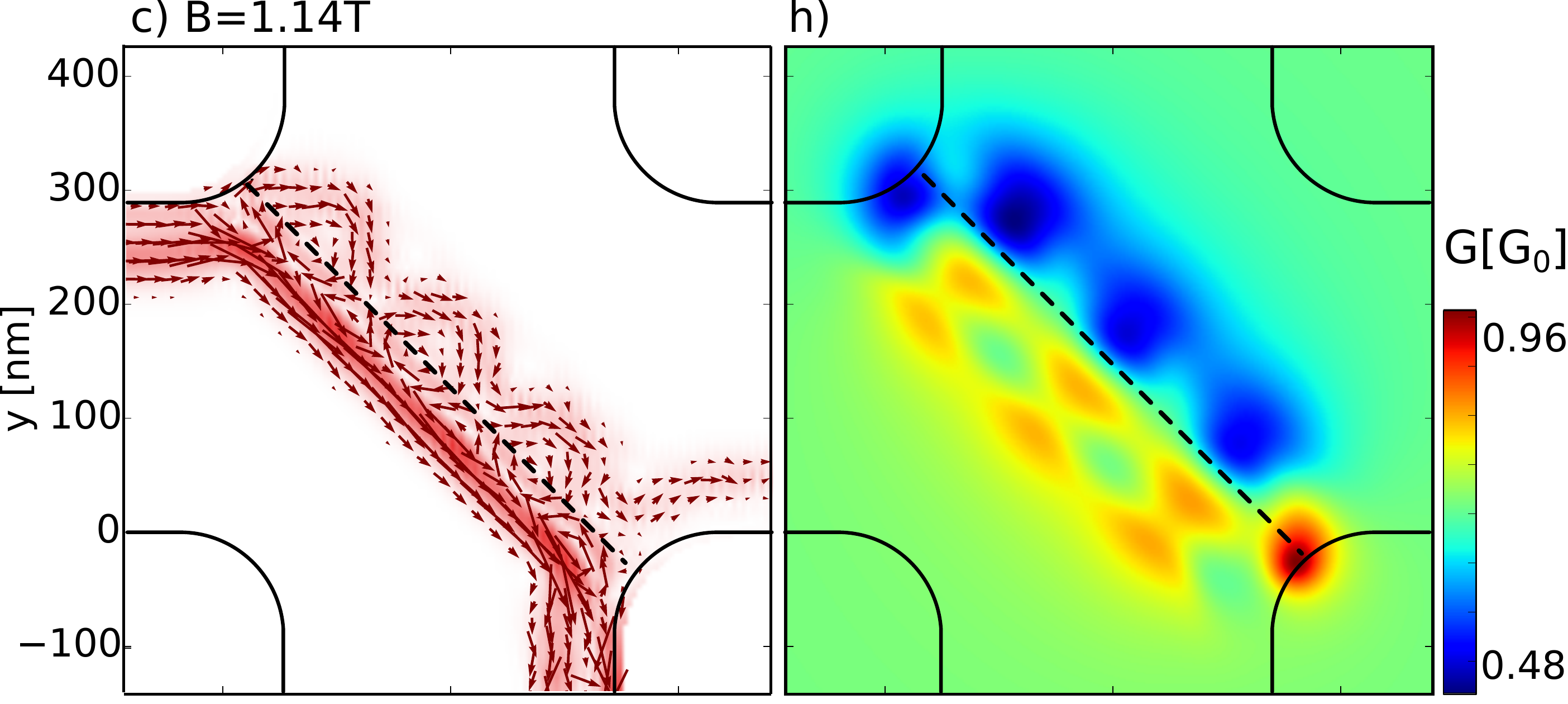} 
\par\end{centering}

\begin{centering}
\includegraphics[width=1\columnwidth]{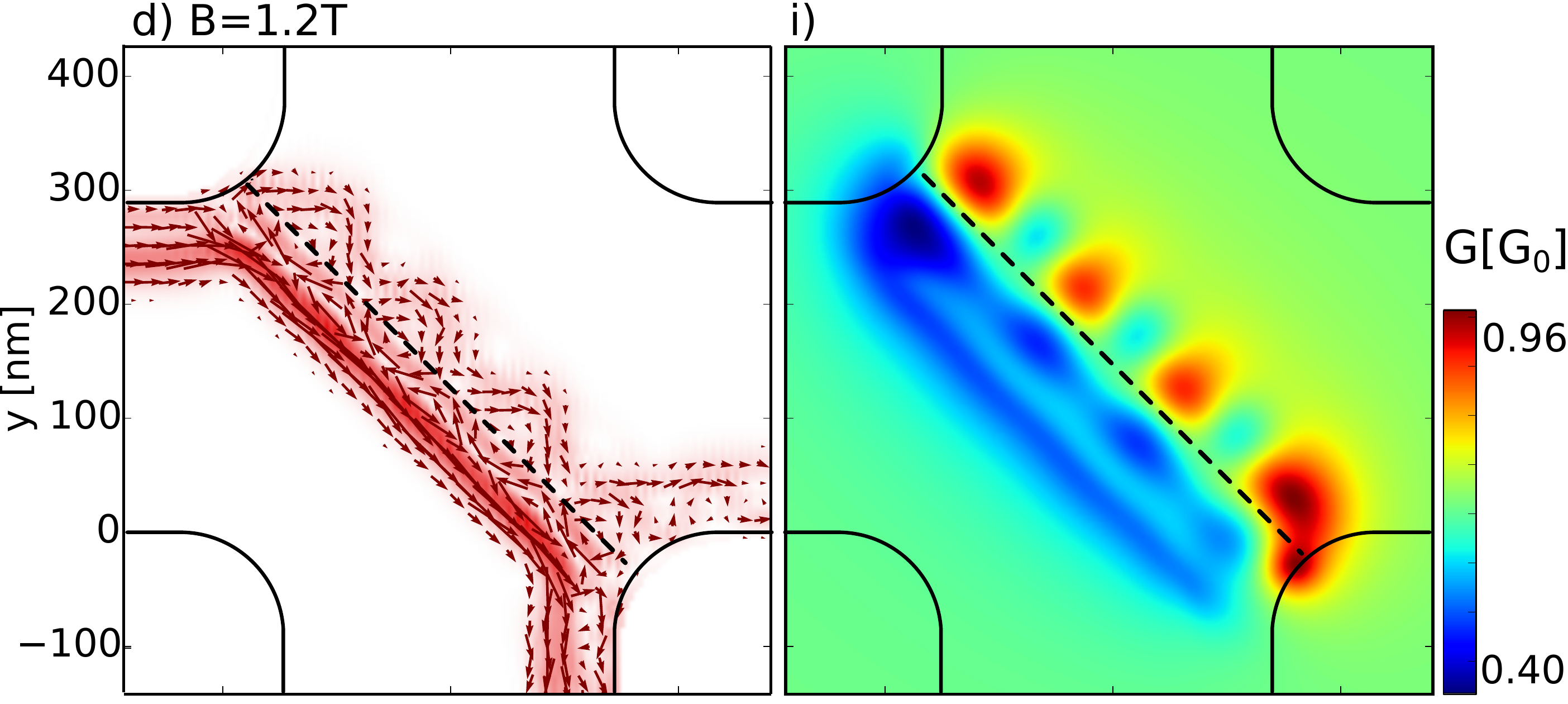} 
\par\end{centering}

\begin{centering}
\includegraphics[width=1\columnwidth]{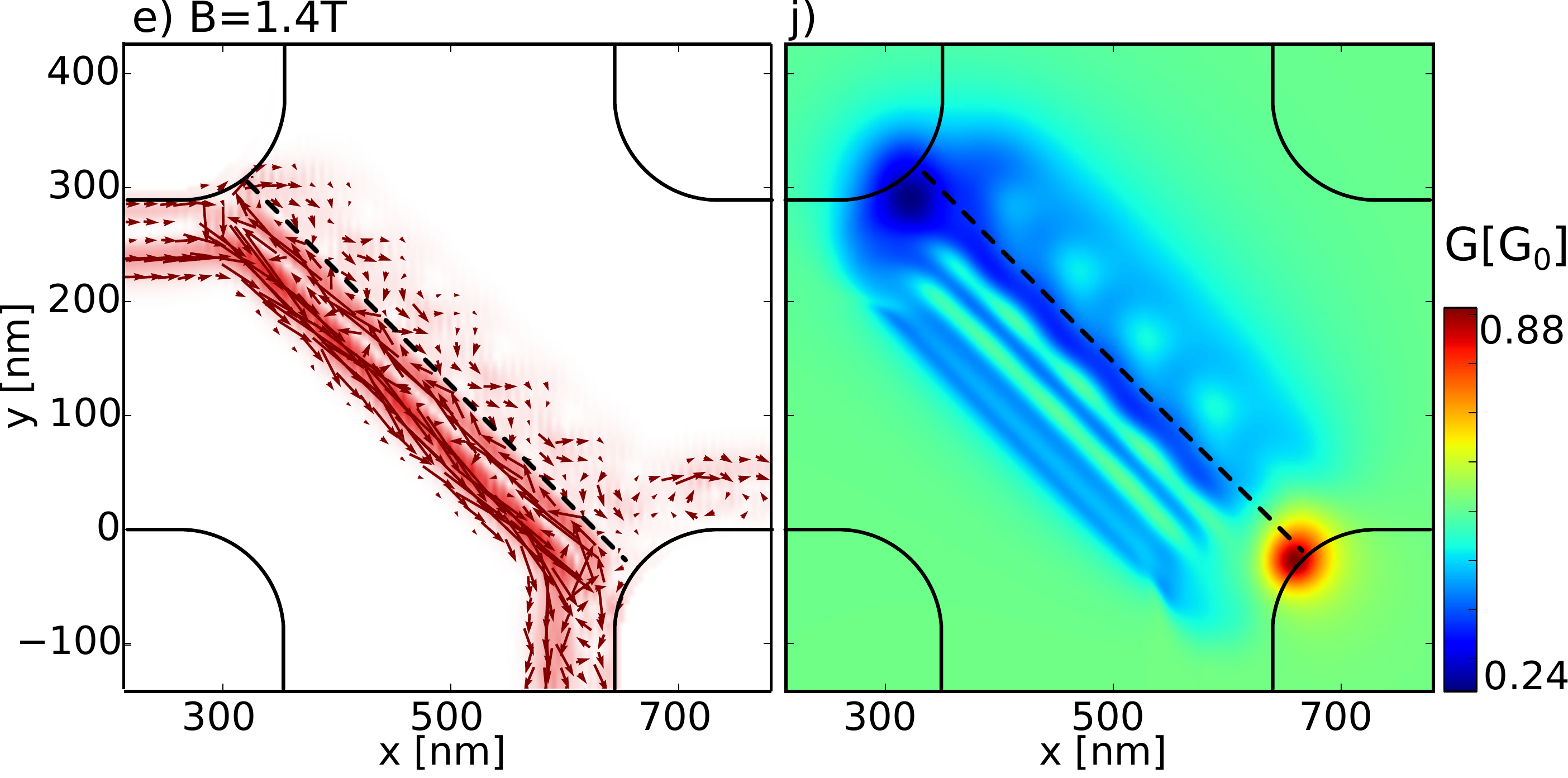} 
\par\end{centering}

\caption{\label{fig:smg}(a-e) Probability current distribution in the absence of the scanning probe
(left column) and (f-j) SGM images (right column) for
magnetic fields $B=0.6$, 1, 1.14, 1.2, 1.4T that are indicated in Fig. \ref{fig:a1}(c).}
\end{figure}

Classically, these oscillations can be understood as due to the variation
of the cyclotron orbits {[}with radius $R_{c}=\hbar k/eB${]} as a
function of $B$. The electron current can be then sent to either
$L_{2}$ or $L_{4}$ lead, depending on the value of $R_{c}$. Figs.
\ref{fig:smg}(a-e) show the electron current distribution for the
electron incident from the lead $L_{1}$ at $E_{F}=40$ meV. The values
of the external magnetic field that are denoted by the corresponding
letters in Fig. \ref{fig:a1}(c). Besides the current confinement
along the n-p junction one notices deflection of the electron paths,
in particular in the p-type (upper right) region of the device, with
a radius that decreases with the external magnetic field.

In terms of the quantum transport theory, the snake orbits along the
n-p junction -- similarly as the skipping orbits \cite{Stegmann2015}
near the edge of the sample -- appear as a result of superposition of
Landau level states 
\begin{equation}
S(x,y)=\sum_{m=1}^{M_{L}}a_{m}e^{ik_{m}x}\chi_{m}(y),\label{eq:S}
\end{equation}
where $k_{m}$ and $\chi_{m}$ denote the $m$th Fermi level wave
vector and the corresponding transverse mode, respectively. The scattering
amplitudes $a_{m}$ depend on specific boundary conditions.

For simplicity let us assume that the particle is propagating along
the $x$ direction. For the case when $M_{L}=1$ the electron density
$\left|S(x,y)\right|^{2}=\left|a_{1}\chi_{1}(y)\right|^{2}$ does
not depend on $x$ and there is no room for the density variation
along the edge or the junction. In this case the idea of the classical
cyclotron orbit is irrelevant. For superposition of two or more modes
the $\left|S(x,y)\right|^{2}$ density results from a superposition
of modes with different Fermi wave vectors, leading to the oscillating
pattern.

\begin{figure}
\begin{centering}
\includegraphics[width=1\columnwidth]{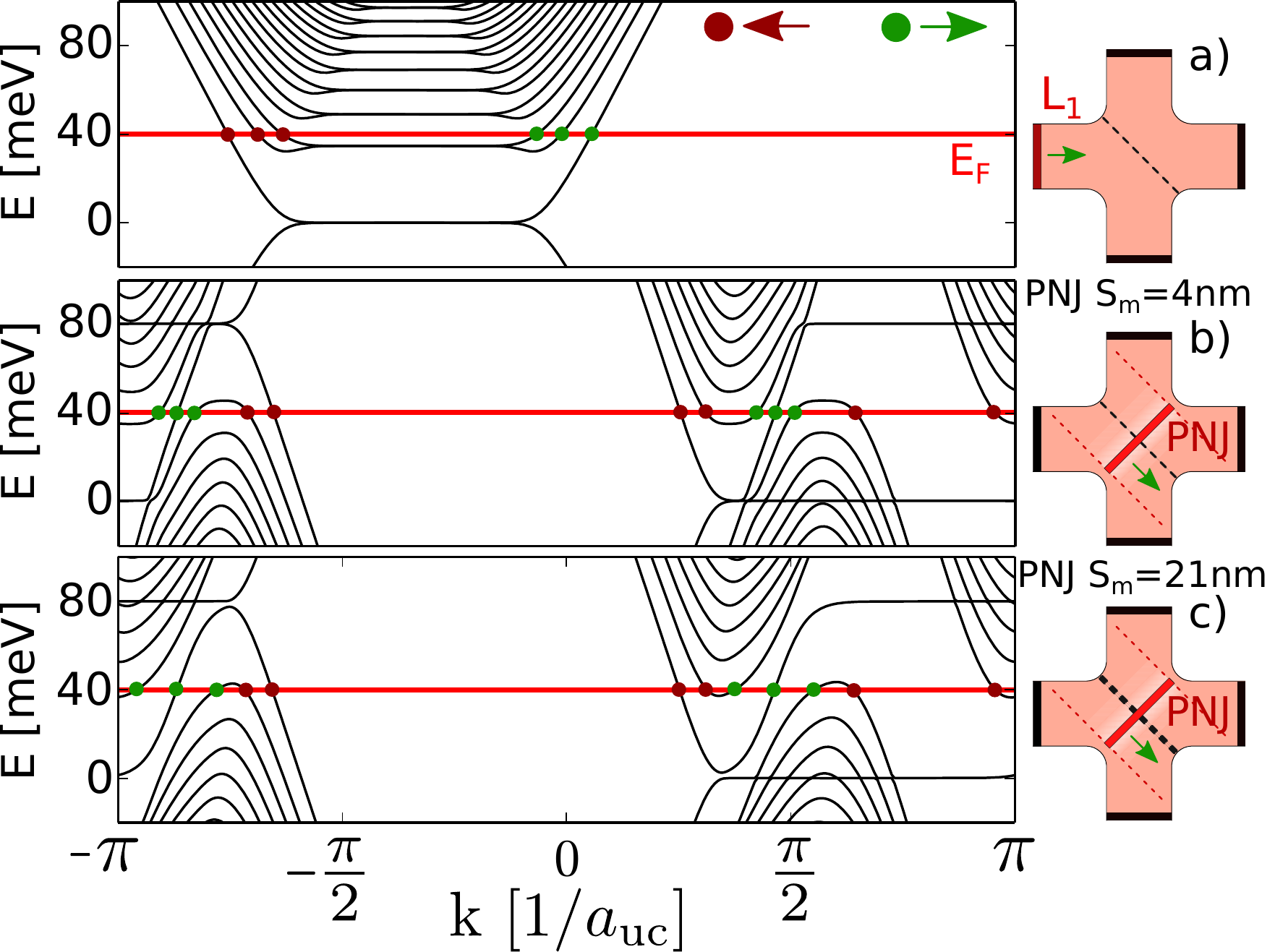} 
\par\end{centering}

\caption{\label{fig:bands}(a) Electron dispersion relation for the lead $L_{1}$. The green and red dots denote the Fermi
level wave vectors for the right- and the left- moving modes, respectively.
(b) Same as (a) but computed for a channel along the n-p junction interface
and for $S_{\mathrm{m}}=4$ nm. (c) same as (b) but for a smoother junction
$S_{\mathrm{m}}=21$ nm. $a_{\mathrm{uc}}$ is the length of the unit
cell vector along which the dispersions are computed. The results were calculated for $B=1.2$ T and $E_{\mathrm{F}}=40$meV.}
\end{figure}

Figure \ref{fig:bands}(a) shows the dispersion relation for a horizontal
channel in lead $L_{1}$ which is denoted by the green arrow in the
inset. For $B=1.2$ T there are three modes which are propagating
to the right (green dots) and three modes for opposite direction (red
dots). The wave vectors $k_{i}$ can be computed from condition $E_{i}(k_{i})=E_{\mathrm{F}}$
\cite{Sorensen2009,Zwierzycki2008}. In Fig. \ref{fig:bands}(b) we
show dispersion relation for a channel created along the n-p junction
of width $S_{\mathrm{m}}=4$ nm (see the inset). For the junction
the number of right-propagating Landau levels is now doubled (six
right and six left propagating modes). The doubling of the states
in the n-p junction was recently discussed for step-like junctions
in Ref. \cite{Mele2015SnakeSym} with reference to the Bogoliubov
quasiparticles and can be explained as a result of coupling the n-type
and the p-type conductivity regions -- in this case with three conduction
and three valence bands at each side of the junction.

\begin{figure}
\begin{centering}
\includegraphics[width=1\columnwidth]{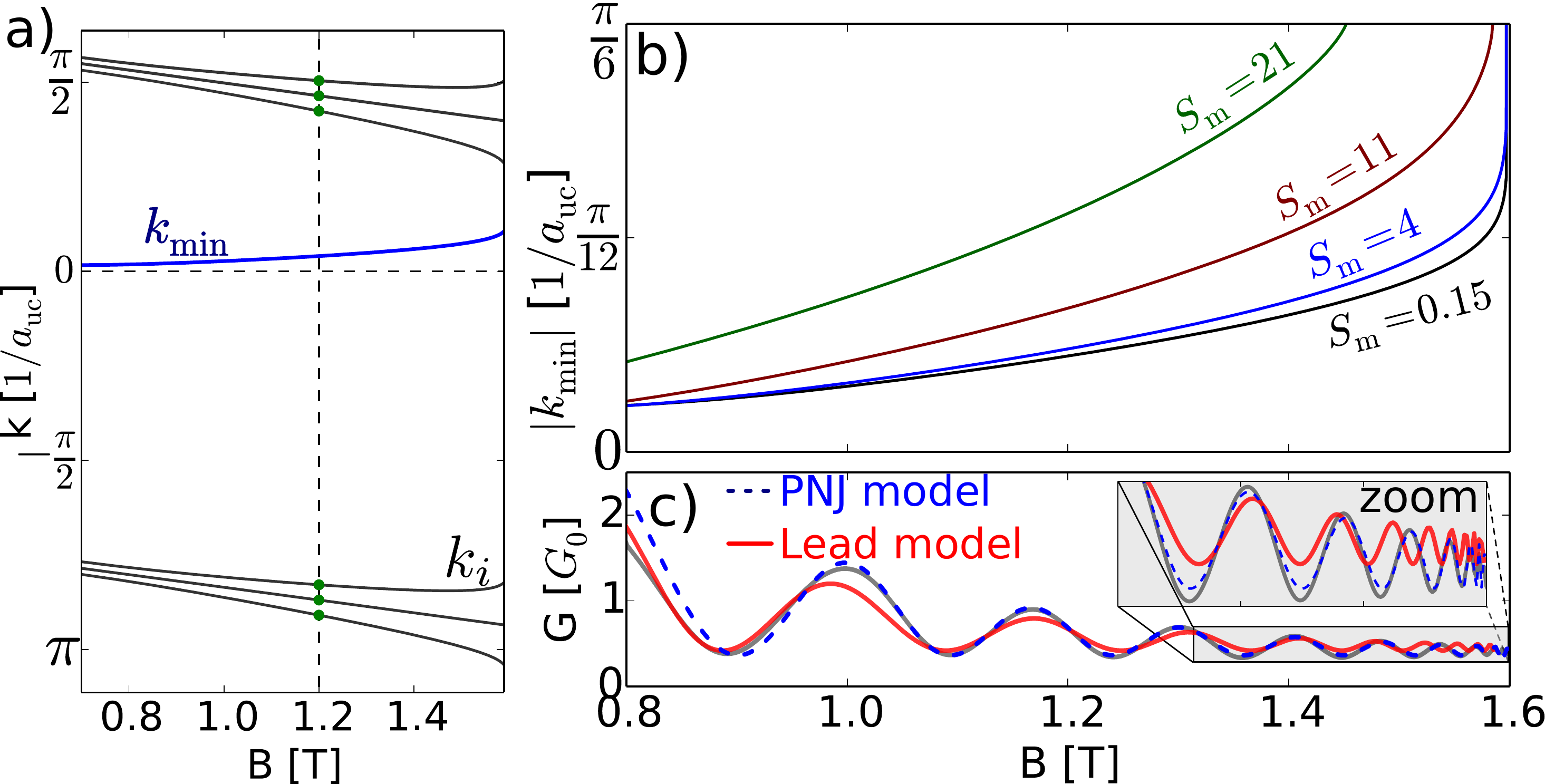} 
\par\end{centering}

\caption{\label{fig:osc}(a) Black lines show the 
the wavevectors $k_{i}$ for the right-going modes at the Fermi level of the n-p junction
interface from Fig. \ref{fig:bands}(b). The vertical dashed line corresponds to  Fig.\ref{fig:bands}(b)
-- see the green dots in both the plots. The blue line show the calculated $k_{\mathrm{min}}$
values according to Eq. (5). (b) $k_{\mathrm{min}}\left(B\right)$ [see Eq. (5)] for different
values of the junction smoothness parameter $S_{\mathrm{m}}$. (c) The gray solid
line the exact result,
the blue dashed line  (PNJ model) -- the best fit obtained from formula \eqref{eq:GBK} 
for the disperion relation of the n-p junction as a conducting channel,
the red line -- the best fit obtained  for the wave vectors of the energy bands of the input lead $L_{1}$.}
\end{figure}

We find that the state doubling is not always present in smooth junctions
of a finite width -- see e.g. Fig. \ref{fig:bands}(b-c) where a small
shift of Fermi energy line will decrease the number of right-moving
modes from 6 to 4. % The reduction of the doubling states%can be easily explained if we consider a smooth transition from step%to smooth interface by taking $S_{\mathrm{m}}\rightarrow+\infty$%which gives a simple channel, hence no doubling.Note,
The slope of the energy bands as function of
the width of the junction $S_{m}$ changes: the plateaux at the extrema of
the bands get narrower in $k$ (c.f. the bottoms of the bands in Fig.
\ref{fig:bands}(b) and (c)) for increased $S_{m}$, which results
in particular in a stronger dependence of the Fermi wave vectors on
both $E_{\mathrm{F}}$ and $B$ off the band extrema. In Fig. \ref{fig:osc}(a)
we show the evolution of the first six Fermi $k_{i}$ wave vectors
for right-moving modes as a function of magnetic field. The black
lines correspond to $S_{\mathrm{m}}=4$ nm. 
%The vertical dashed line marks $B=1.2$ T assumed in Fig. \ref{fig:bands}(b). 

The conductance oscillations of Fig. \ref{fig:a1}(c) can be explained
by the superposition of the modes propagating along the junction.
We consider the Fermi wave vectors and look for the closest pair of
$k$'s that correspond to the positive velocity $dE/dk>0$. A superposition
{[}see Eq. \eqref{eq:S}{]} of these two modes produces a charge density
variation of the largest wavelength $\lambda_{\mathrm{max}}=2\pi/k_{\mathrm{min}}$,
\begin{equation}
k_{\mathrm{min}}\equiv\min_{i,j}\left(k_{i}-k_{j}\right).\label{eq:Kmin}
\end{equation}
Within the range of the magnetic field from $B=0.8$ T to 1.6 T the
minimal distance between the right-going wave vectors appears for
the two lowest ones in Fig. \ref{fig:osc}(a) and the two highest
ones. Both distances are found equal. For a superposition of
the two modes with the closest wave vectors $S=a_{i}\exp(ik_{i}x)\chi_{1}(y)+a_{j}\exp(ik_{j}x)\chi_{2}(y)$.
Hence the charge density along the junction can be put in a form $|S|^{2}=|a_{i}|^{2}|\chi_{i}(y)|^{2}+|a_{j}|^{2}|\chi_{j}(y)|^{2}+2\Re(a_{i}^{*}a_{j})\chi_{i}^{*}(y)\chi_{j}(y)-4\Re(a_{1}^{*}a_{2})\sin^{2}\left(\frac{(k_{i}-k_{j})x}{2}\right)\chi_{i}^{*}(y)\chi_{j}(y)$.
The last term is responsible for the oscillations of the density along
the junction, and the $G_{21}$ conductance depends on the destination
of the current that reaches the end of the n-p junction at its contact
with the edge. At that point the electron current reaches either $L_{4}$
or $L_{2}$. The oscillations of Fig. \ref{fig:a1}(c) can be described
by a a simple phenomenological formula 
\begin{equation}
G(B)=a+be^{-cB}\sin^{2}\left(\frac{k_{\mathrm{min}}\left(B\right)L}{2}\right),\label{eq:GBK}
\end{equation}
where the exponential part accounts for the observed decay of the
oscillations at high $B$ and $L=455$ nm is the length of the junction.

The decrease of the amplitude for higher magnetic field can be understood
based on the recent Ref. \cite{szaczen}, which shows that the transmission
coefficient for electron traveling through the n-p junction of a finite
width is below 1. Here we consider the junction with $S_{\mathrm{m}}=4$ nm
which according to Ref. \cite{szaczen} gives the transmission probability
$\sim0.5$. Now, if we increase the magnetic field the number of times
that electron passes across the n-p interface increases, hence the
reduction of the conductance oscillations amplitude.

The results of the fit with formula \eqref{eq:GBK} for the dispersion
relation for the n-p junction {[}Fig. \ref{fig:bands}(b){]} are denoted
as the ''PNJ model'' in Fig. \ref{fig:osc}(c) -- the dashed blue line,
which above 0.9T agrees perfectly with the numerical conductance.
A similar analysis was performed in Ref. \cite{Milovanovic2014} but
for the dispersion relation of the input lead and not the n-p junction itself. The fit for the dispersion
relation of the input lead is given by the red line in Fig. \ref{fig:osc}(c),
in which the agreement is not as good as for the ''PNJ model''. Note, that the fit becomes even
worse for a larger magnetic fields, where the out of phase range is visible (see zoomed area in Fig.\ref{fig:osc}(c)).
A distinct shift can also be spotted in Fig. 4(f) of Ref. \cite{Milovanovic2014}
between the model and numerical values, however at lower $B$. The present
result indicate that the properties of the band structure of the n-p
junction as the conducting channel precisely determine the period
of the conductance oscillations on the magnetic field scale in the
quantum Hall conditions.

\subsection{Imaging the snake states}

\subsubsection{Weak perturbation}

As  weak perturbation we consider the tip potential of $V_{\mathrm{tip}}=10$meV -- four times smaller
than the Fermi energy.
In Figs. \ref{fig:smg}(f-i) we show the SGM conductance maps for
the current distribution given in the left column in Fig. \ref{fig:smg}(a-e).

The conductance does react to the external perturbation
-- but only for the scanning probe near the n-p junction.
No effect is observed for the tip in the leads. The
tip deflects the trajectory to $L_{4}$ or $L_{2}$ leads
but no backscattering is present which is characteristic to the quantum Hall conditions,
hence the flat maps for the probe above the leads. 

Figures \ref{fig:smg}(a,f) correspond to the magnetic field where
a conductance peak {[}Fig. \ref{fig:osc}(c){]} is observed. For this
magnetic field the cyclotron radius is comparable to the length of
the n-p junction. The SGM image {[}Fig. \ref{fig:a1}(f){]} does
not resolve the details of this orbit. Moreover, here and for other
$B$ values the SGM maps have an approximate symmetry with respect
to the inversion through the bisector of the junction (here $y=x$
line) which is missing in the current plots.

For the subsequent conductance peak marked by "b)" in Fig. 2(c) 
  the cyclotron radius of the deflected electron trajectory
{[}Fig. \ref{fig:osc}(c){]}
 is already $\simeq5$ times shorter than $L$ {[}see
Fig. \ref{fig:smg}(b){]} and the distance between the extrema of
conductance map {[}Fig. \ref{fig:smg}(g){]} along the junction is
comparable to the cyclotron radius. This also found for higher magnetic
fields -- Fig. \ref{fig:smg}(c,h), \ref{fig:smg}(d,i) and \ref{fig:smg}(e,j),
although the visibility of the oscillation becomes unequal at the opposite	
sides of the junction. The non-transparency of the finite-width junction
for electrons \cite{szaczen}, discussed in the context of Fig. \ref{fig:osc}(c)
is one of the possible reasons responsible for the reduction of the
conductance visibility at high magnetic field.

\subsubsection{Strong perturbation}

\begin{figure}
\begin{centering}
\includegraphics[width=1\columnwidth]{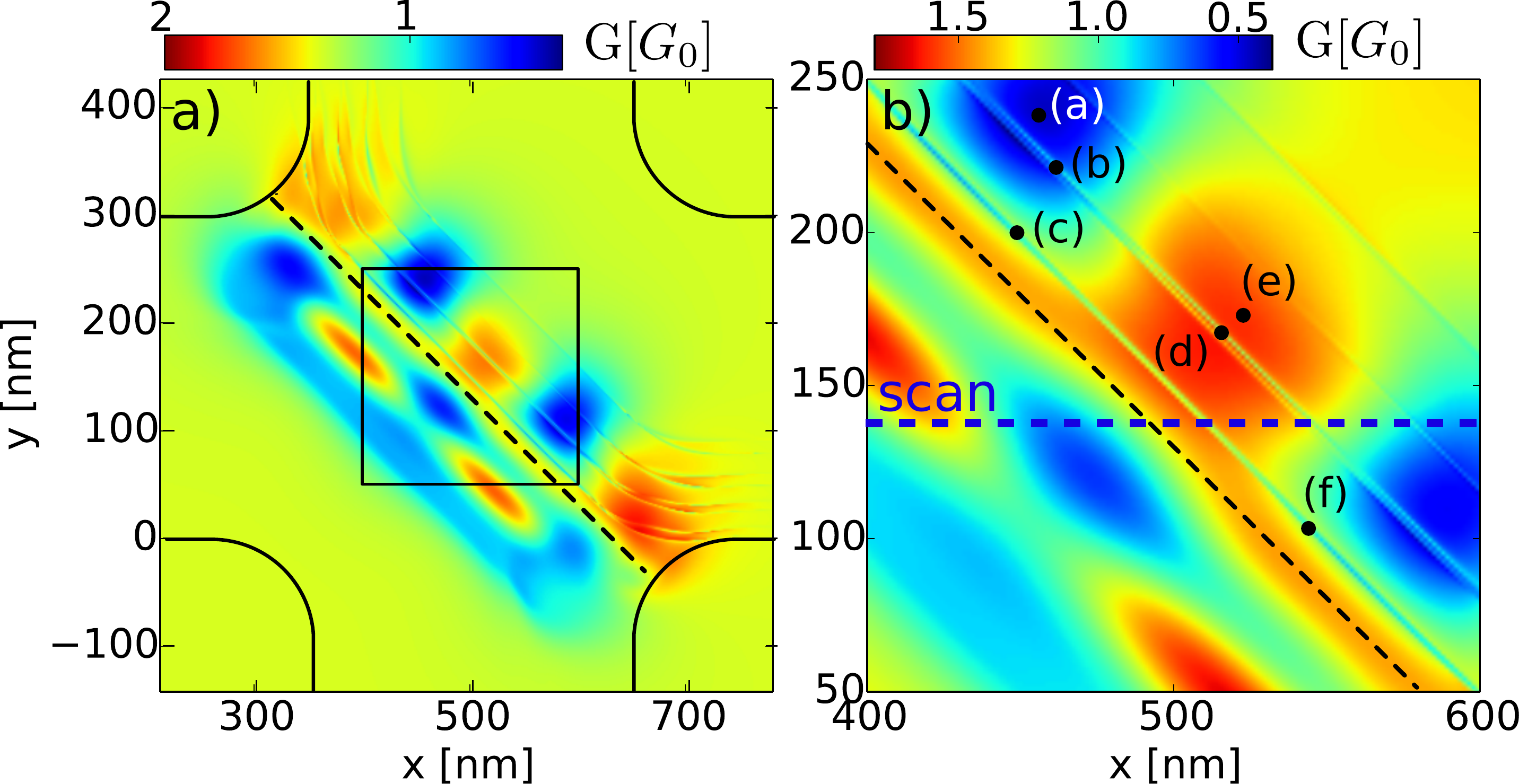} 
\par\end{centering}

\caption{\label{fig:strong}(a) The conductance map for $V_{\mathrm{tip}}=30$meV
and $B=1$ T. (b) The zoom of the black rectangle in (a). The labels correspond
to tip locations considered in Fig. \ref{fig:cd}. %For the horizontal dashed line
%the conductance map was given in Fig. \ref{szkan}.
}
\end{figure}

\begin{figure}
\begin{centering}
\includegraphics[width=1\columnwidth]{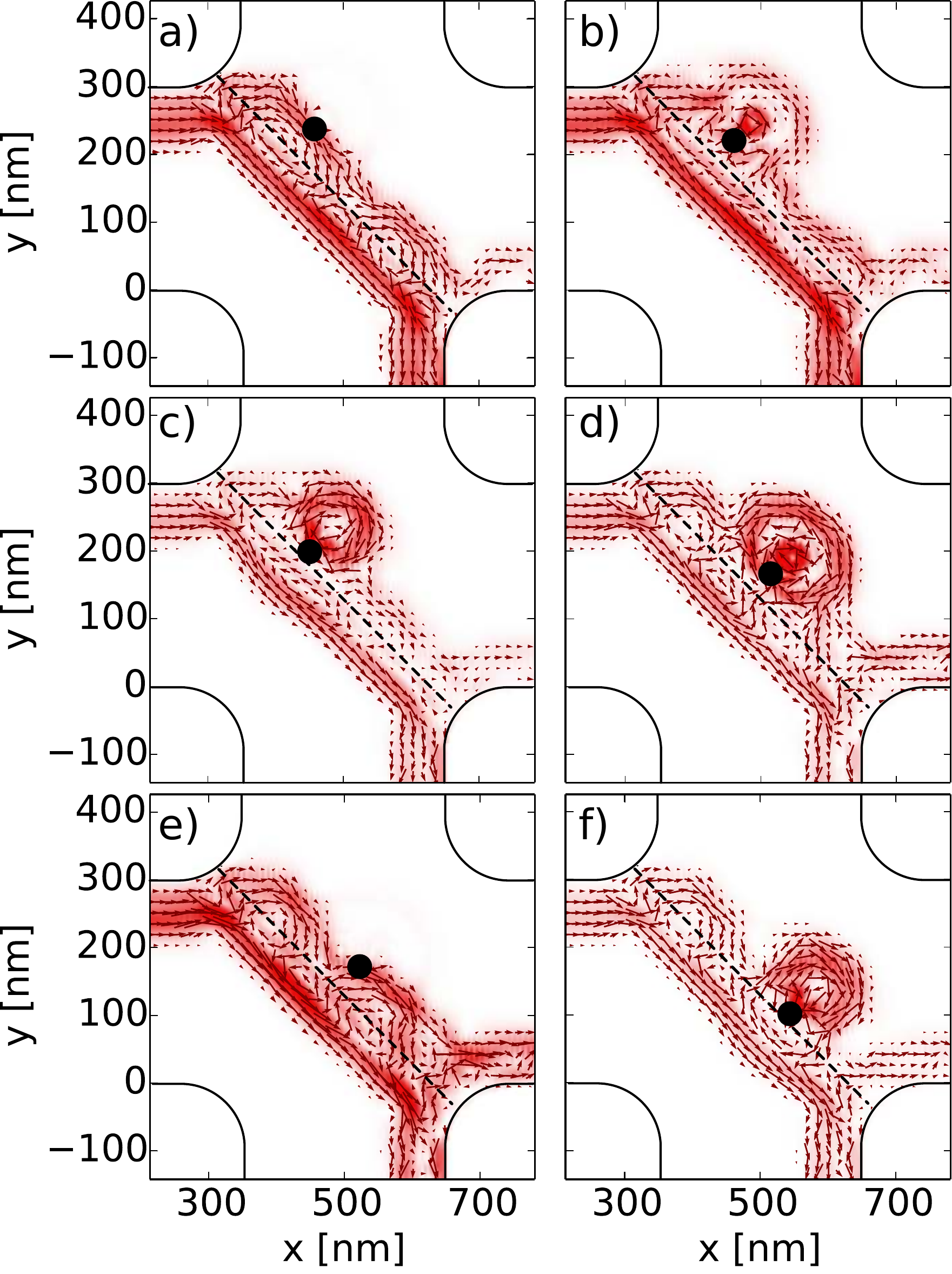} 
\par\end{centering}

\caption{\label{fig:cd} Probability current distribution obtained for the probe
locations denoted in 
Fig. \ref{fig:strong}(b).}
\end{figure}

\begin{figure}
\begin{centering}
\includegraphics[width=1\columnwidth]{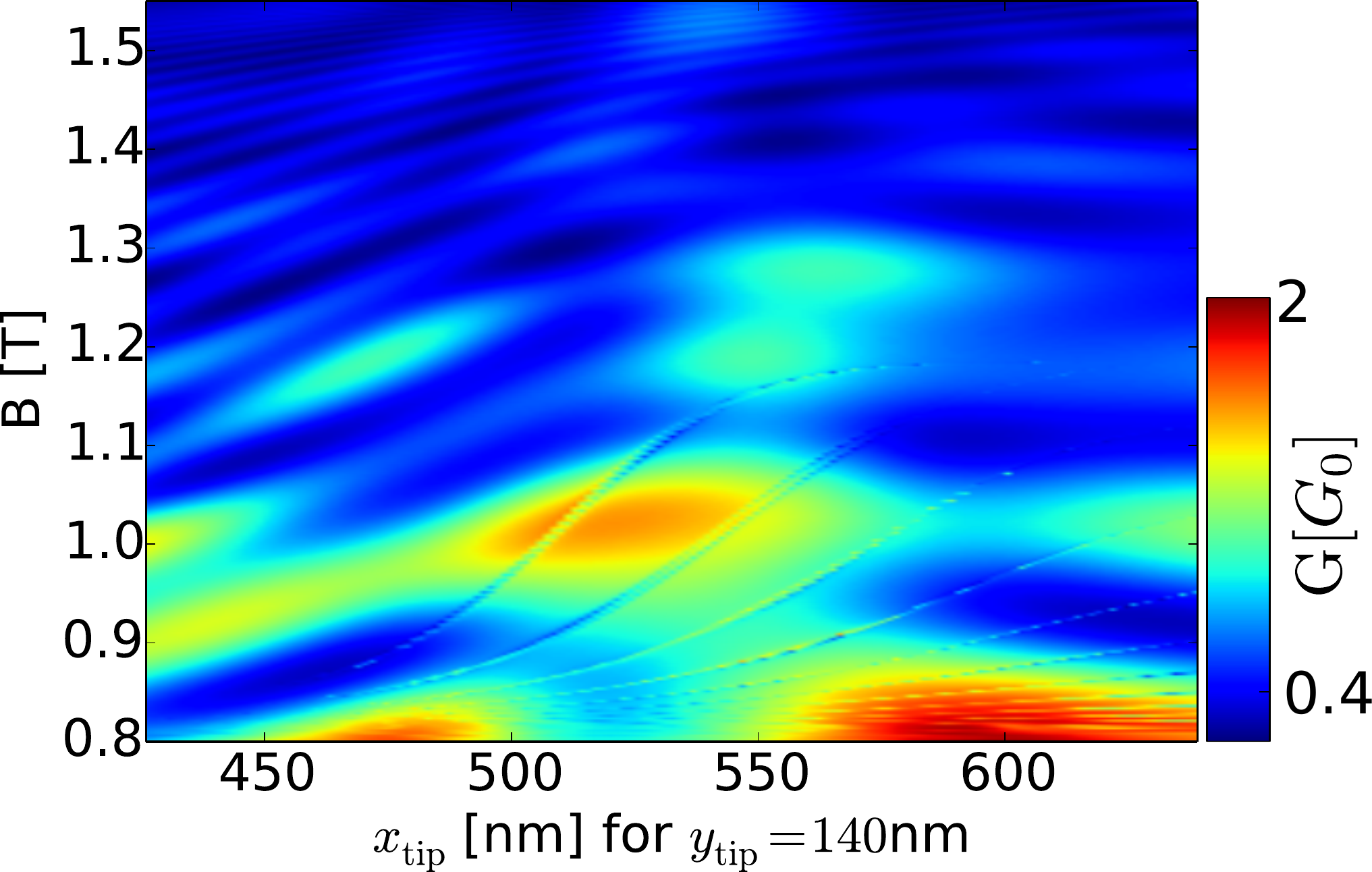} 
\par\end{centering}

\caption{\label{szkan} The conductance along the horizontal dashed line in Fig. \ref{fig:strong}(b)
as a function of the magnetic field.}
\end{figure}

The SGM images for a stronger tip potential $V_{\mathrm{tip}}=30$
meV and $B=1$ T are displayed in Fig. \ref{fig:strong} -- to be
compared with Fig. \ref{fig:smg}(b) for $V_{\mathrm{tip}}=10$ meV.
The asymmetry of the plot between the n and p sides of the junction
is increased for larger $V_{\mathrm{tip}}$. Moreover, a number of
resonances is found at the p side at lines parallel to the junction.
The current density plots for tip location over the points indicated
in Fig. \ref{fig:strong}(b) are displayed in Fig. \ref{fig:cd}.
The resonances are related to current loops that encircle the tip.
The current loops are found for the tip on the p side only. The
tip potential repels the carriers on the conduction band side of the junction. The
currents on the n-side simply avoid the perturbation and no loop of
current is found. For the carriers on the valence band side the potential
maximum induced by $V_{tip}$ is attractive and thus it supports a
quasi-bound state. The exact positions of the resonances depend on
the magnetic field in a periodic manner -- which is related to the
Aharonov-Bohm effect for the current circulation around the tip that
couples to the junction current. The conductance across the junction
is displayed in Fig. \ref{szkan} along the dashed line marked in
Fig. \ref{fig:strong}(a) as a function of the external magnetic field.
For lower magnetic fields the resonances are found also for the tip
near the n-p junction on the n side. 
 For a given tip location the
spacing between the subsequent resonances depends on the magnetic
field. For higher magnetic fields the clockwise loop that is
seen in Fig. \ref{fig:cd} is made tighter by the Lorentz force which
acts to the right of the current orientation on the p conductivity
side. For a reduced radius of the current loop the magnetic field
period corresponding to a flux quantum is increased.

\subsection{Conductance maps for wider n-p junctions}

\begin{figure*}[!t]
\begin{centering}
\includegraphics[width=1\textwidth]{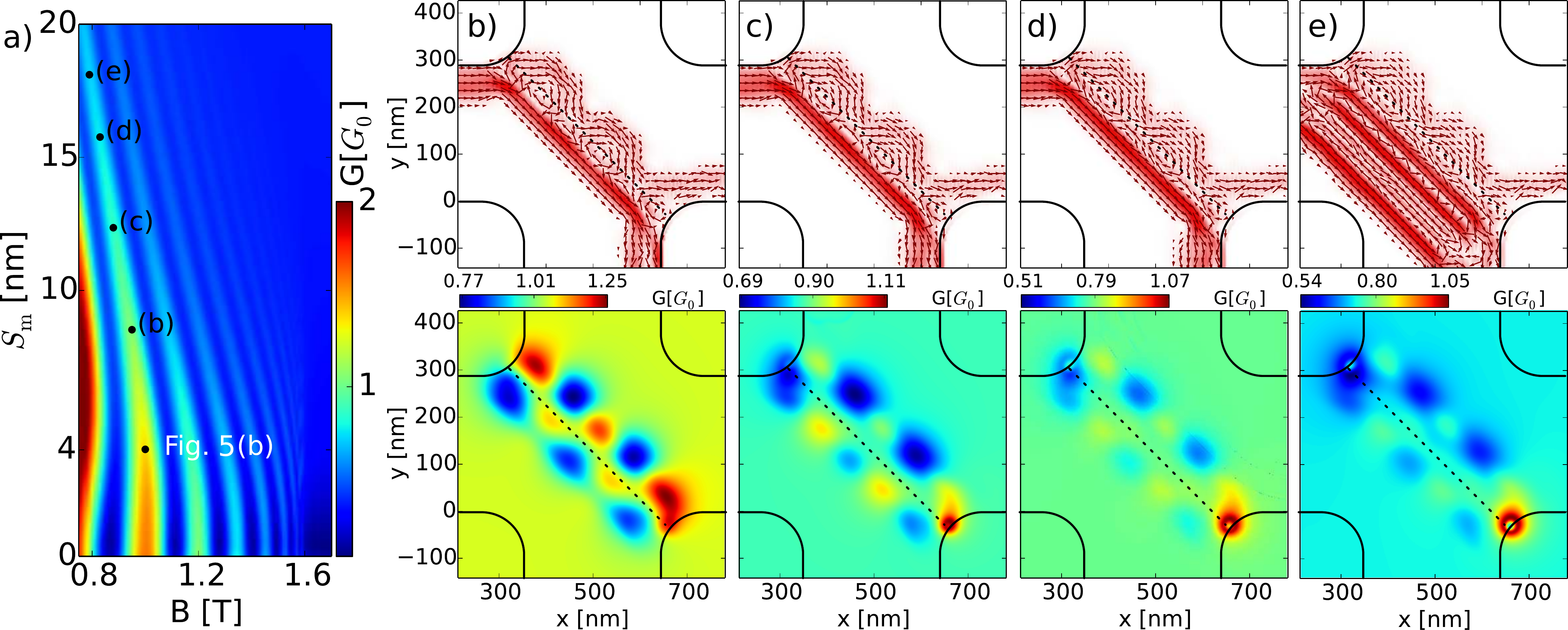} 
\par\end{centering}

\caption{\label{fig:sm}(a) The conductance as a function of magnetic field and
the junction smoothness parameter $S_{\mathrm{m}}$. The white label corresponds to
the case shown in Fig. 5(b). (b-e) The probability current distribution and
the corresponding conductance maps for the work points marked in  (a).}
\end{figure*}

So far we have discussed the case of a thin junction with $S_{\mathrm{m}}=4$
nm. From the discussion in Section III.A we know that the width of the
junction strongly affects the dispersion relation. We calculated the
conductance at $E_{F}=40$ meV (as in Fig. \ref{fig:a1}(c)) as a
function of the junction width $S_{\mathrm{m}}$ and the magnetic
field. In the result presented in Fig. \ref{fig:sm}(a) one notices
that (i) the resonance lines bend towards lower magnetic field as
$S_{\mathrm{m}}$ is increased and (ii) the amplitude of the oscillations
decreases with $S_{\mathrm{m}}$. The feature (i) results from the
fact that the spacing between the nearest $k$ vectors is increased
for wider junctions and at higher magnetic field {[}Fig. \ref{fig:osc}(b){]}.
The corresponding resonances appear for smaller magnetic field values
at larger $S_{m}$. The finding (ii) seems due to a decreased transparency
of the junction with its width found recently in Ref. \cite{szaczen}.
To summarize, we find that for a smooth junction snake orbits appear
for lower magnetic fields but at the expense of the visibility of
oscillations.

Figures \ref{fig:sm}(b-e) show the probability current plots and
SGM conductance maps for the work points marked in Fig. \ref{fig:sm}(a)
and $V_{\mathrm{tip}}=10$meV along a selected resonance line. One
may see that once we increase $S_{\mathrm{m}}$ the snake features
become less resolved both in the current plots and the conductance
maps. The current along the n side increases with $S_{\mathrm{m}}$.
The deflected trajectories remain at the p side, where also the
amplitude of the conductance map remain stronger than on the n
side. In Fig. \ref{fig:sm}(e) the number of spin degenerated modes
is 5 instead of 3, hence the current plot contains additional features
from higher modes. However, in the corresponding SGM image no additional
features are found.

\section{Summary and conclusions}

We have discussed the conductance mapping of the snake orbits confined
at the n-p junction in graphene. We indicated a precise relation of
the conductance oscillations at the magnetic field scale with the
Fermi wavelengths of the n-p junction as a waveguide. We found that
the maps of conductance contain oscillating patterns along the
junction with the period of the oscillation that is close to the period
of the snake orbit. The visibility of the map decreases with the external
magnetic field due an increased number of the electron passages across
the junction and a non-ideal transparency of the n-p junction of a
finite width. The conductance maps are found to be nearly symmetrical
across the bisector of the junction with an asymmetry between the
n- and p- sides. For stronger tip potentials resonant quasi-bound
states are formed under the tip at one of the junction sides. The
interference of the quasi-bound states with the junction currents
produces resonances parallel to the junction with positions that react
strongly to the external magnetic field via  the Aharonov-Bohm phase shift.
We demonstrated that the width of the n-p interface affects the oscillation
period and the visibility of the conductance maps.

\section*{ Acknowledgments} 
This work was supported by the National Science Centre (NCN) according to decision DEC-2015/17/B/ST3/01161.
The calculations were performed on PL-Grid Infrastructure. 

\bibliographystyle{apsrev4-1-nourl}

\end{document}